 \documentclass[10pt]{article}

\usepackage{fullpage}

\usepackage{amsmath,amsfonts,amssymb}
\usepackage{graphicx}
\usepackage{setspace}
\usepackage{tocloft}
\usepackage{graphicx}
\usepackage{epstopdf}
\usepackage{subcaption}
\usepackage{ulem}
\usepackage{empheq}
\usepackage{color}
\usepackage{adjustbox}
\usepackage{balance}

\usepackage{multirow}

\usepackage{upgreek}

\usepackage{widetext}

\usepackage{cite}

\def\##1{{\bf#1}}
\def\=#1{\underline{\underline{#1}}}

\def\+#1{\underline{\bf #1}}
\def\*#1{\underline{\underline{\bf #1}}}

\def\les{\left[}
\def\ris{\right]}
\def\lec{\left\{}
\def\ric{\right\}}
\def\lek{[{\kern 0.1em}}
\def\rik{{\kern 0.1em}]}

\def\.{\mbox{ \tiny{$^\bullet$} }}

\def\.{\mbox{ \tiny{$^\bullet$} }}

\def\eps{\varepsilon}

\def\lambdao{\lambda_{\scriptstyle 0}}

\def\degC{$^\circ$C}

\def\vo{VO$_2$}
\def\epsvo{\eps_{\rm vo}}
\def\epsr{\eps_{{\rm vo}}}

\def\epsvom{\eps_{\rm vo}^{\rm mono}}
\def\epsvot{\eps_{\rm vo}^{\rm tetra}}

\def\nr{n_{{\rm vo}}}
\def\nrm{n_{\rm vo}^{\rm mono}}
\def\nrt{n_{\rm vo}^{\rm tetra}}

\def\bara{\bar{a}}

\def\sca{_{\rm sca}}
\def\forw{_{\rm f}}
\def\back{_{\rm b}}
\def\abs{_{\rm abs}}
\def\ext{_{\rm ext}}
\def\pr{_{\rm pr}}

\def\DQ{\Delta{Q}}

\def\one{^{(1)}}

\def\three{^{(3)}}

\begin{document}
  
  \begin{center}
\textbf{Thermal hysteresis in scattering by vanadium-dioxide spheres }\\
\vspace{3mm}

{Akhlesh Lakhtakia}$^{1,*}$, {Tom G. Mackay}$^{1,2}$, and
{Waleed I. Waseer}$^{1,3}$\\
\vspace{3mm}

$^1$ {The Pennsylvania State University, Department of Engineering Science and Mechanics, University Park, Pennsylvania 16802, United States of America}\\
\vspace{3mm}
$^2$ {University of Edinburgh, School of Mathematics and Maxwell Institute for Mathematical Sciences, Edinburgh EH9 3FD, Scotland, UK}\\
\vspace{3mm}
$^3$ {COMSATS University Islamabad, Department of Electrical and Computer Engineering, Islamabad, Pakistan}\\
\vspace{3mm}
$^*${Corresponding author: akhlesh@psu.edu}
\vspace{3mm}
\end{center}

\begin{abstract} 
Vanadium dioxide (\vo) transforms from purely monoclinic to purely tetragonal on being heated
 from  58~\degC~to 72~\degC, the transformation being reversible
but hysteretic. Electromagnetically, \vo~transforms from a dissipative dielectric to another dissipative dielectric
if the free-space wavelength  $\lambdao < $  1100~nm, but from a dissipative dielectric to a plasmonic metal (or \textit{vice versa})
if $\lambdao>1100$~nm. Calculating the  extinction, total scattering, absorption, radiation-pressure, back-scattering, and forward-scattering 
 efficiencies of a \vo~sphere, we found
clear signatures
of thermal hysteresis  in (i) the forward-scattering,  back-scattering, and absorption efficiencies for  $\lambdao<1100$~nm,
and (ii) the forward-scattering,   back-scattering, total scattering, and absorption efficiencies for $\lambdao>1100$~nm.
Vacuum and null-permittivity quasistates occur between 58~\degC~and 72~\degC, when tetragonal \vo~is a plasmonic metal,
once each on the heating branch and once each on the cooling branch of thermal hysteresis. But
none of the six efficiencies show significant differences
between the two quasistates.

 \end{abstract}

\section{Introduction}
Vanadium dioxide (\vo) is monoclinic at any temperature $T\lesssim 58$~\degC~but tetragonal 
for $T\gtrsim 72$~\degC. The crystal structure
changes  as the temperature is raised
from below $58$~\degC~to above $72$~\degC~\cite{Morin}, a mixture of crystals of both types existing in the intermediate thermal regime
\cite{Choi}. During the transformation,
the  vanadium--vanadium atomic pairs located periodically in the monoclinic crystal separate
into single  atoms located periodically in the tetragonal crystal \cite{Horlin}, the mass density increases
by 1.79\%, and the electrical conductivity increases by 
at least three orders of magnitude \cite{Morin,Crunteanu}. Not only is the transformation  thermally reversible,
its electromagnetic consequences are clearly evident in the sub-megahertz \cite{Mansingh}, megahertz \cite{Yang},
gigahertz \cite{Crunteanu,Emond}, far-infrared \cite{Gao,Parrott}, mid-infrared \cite{Choi}, near-infrared \cite{Cormier,Skelton}, and visible \cite{Cormier,Kepic} spectral regimes.  Therefore, this material is attractive for a variety of switching applications \cite{Dumas,Wang2015,Hashemi,Sereb2022,Tripathi}.

Monoclinic \vo~is a dissipative insulator, i.e., both  the real and the imaginary parts of
the relative permittivity $\epsvom$ are positive, with ${\rm Re}\lec\epsvom\ric$ significantly larger than ${\rm Im}\lec\epsvom\ric$
in the visible and sub-visible spectral regimes \cite{Kepic,Cormier,Mansingh,Gao,Parrott,Horlin,Choi,Skelton,Yang,Crunteanu,Emond}.
However, tetragonal \vo~is a dissipative insulator only when the free-space wavelength $\lambdao\lesssim1100$~nm \cite{Kepic},
with ${\rm Re}\lec\epsvot\ric>0$ and ${\rm Im}\lec\epsvot\ric>0$. 
But  ${\rm Re}\lec\epsvot\ric<0$ and ${\rm Im}\lec\epsvot\ric>0$ for  $\lambdao\gtrsim1100$~nm, so that tetragonal \vo~then functions as a plasmonic metal. The complete insulator-to-metal transformation (IMT)
on heating by  $14$~\degC~and the complete metal-to-insulator transformation (MIT) on cooling by $14$~\degC,  at reasonably low temperatures
in the neighborhood of   $65$~\degC, are responsible for the attraction of \vo~for switching applications for $\lambdao\gtrsim1100$~nm
\cite{Dumas,Wang2015,Hashemi,Sereb2022,Tripathi}. These applications  
are premised on two unambiguous states: the {\sf ON} state in which \vo~is purely monoclinic/tetragonal
and the {\sf OFF} state  in which \vo~is purely  tetragonal/monoclinic.

Although reversible, the thermal transformation is hysteretic. In other words, the properties of \vo~in the intermediate thermal regime
($58$~\degC~to $72$~\degC) depend on whether the material is being heated/cooled from having only monoclinic/tetragonal
 crystals to  having only
tetragonal/monoclinic  ones. The electromagnetic consequences of thermal hysteresis remain obscure, although
the relative
permittivity $\epsvo(\lambdao,T)$ has been investigated  on both the heating and the cooling branches by a few researchers
\cite{Cormier,Choi,Taylor}.

Scattering by an isotropic dielectric sphere is a basic electromagnetic boundary-value problem \cite{Logan}, with its 
solution traceable to Lorenz \cite{Lorenz1,Lorenz2},
Mie \cite{Mie}, and Debye \cite{Debye}. Since  
$\epsvo(\lambdao,T)$ in the intermediate thermal regime is bounded by $\epsvom(\lambdao)$ and $\epsvot(\lambdao)$ \cite{Cormier,Choi,Taylor},
and $\epsvom(\lambdao)$ differs from $\epsvot(\lambdao)$,  the scattering characteristics of a \vo~sphere of radius $a$ at any specific value of
$\lambdao$
will depend not only on the ratio $\bara=a/\lambdao$ but also on $T$. If these characteristics on the heating branch are
sufficiently different from those on the cooling branch, they could be useful in determining not only the ambient temperature 
$T\in[58\,^\circ{\rm C},72\,^\circ{\rm C}]$
but also
 whether the surroundings of a \vo~sphere are being heated or cooled in the intermediate thermal regime. 

That idea motivated this research. Standard expressions devolving from the Lorenz--Mie theory 
for plane-wave scattering by an isotropic dielectric sphere \cite{BH83}  were
used for the extinction, total scattering, absorption, radiation-pressure, back-scattering, and forward-scattering 
 efficiencies of a \vo~sphere. These quantities were computed as functions
of temperature in the intermediate thermal regime and the manifestations of thermal hysteresis   in
electromagnetic scattering were identified.

\section{Theoretical expressions}
Suppose that a plane wave propagating in a fixed direction illuminates the sphere $r\leq a$. Then,
\begin{equation}
Q\ext = \frac{1}{2\pi^2\bara^2}{\rm Re}\lec\sum_{n=1}^{\infty}
\les(2n+1) \left(a_n+b_n\right)\ris\ric\,
\end{equation}
is the extinction efficiency,
\begin{equation}
Q\sca= \frac{1}{2\pi^2\bara^2} \sum_{n=1}^{\infty}
\les(2n+1) \left(\vert{a_n}\vert^2+\vert{b_n}\vert^2\right)\ris\,
\end{equation}
is the total scattering efficiency,
\begin{equation}
Q\abs=Q\ext -Q\sca\,
\end{equation}
is the absorption efficiency,
\begin{eqnarray}
&&
\nonumber
Q\pr= Q\ext-\frac{1}{\pi^2\bara^2}{\rm Re}\lec
\sum_{n=1}^{\infty}
\les
\frac{n(n+2)}{n+1}\left(a_n^\ast a_{n+1} + b_n^\ast b_{n+1}\right)
\right.\right.
\\[5pt]
&&\left.\left.
\qquad\qquad+
\frac{2n+1}{n(n+1)}a_n^\ast b_n
\ris
\ric
\end{eqnarray}
is the radiation-pressure efficiency with the asterisk
denoting the complex conjugate,
\begin{equation}
Q\back= \frac{1}{4\pi^2\bara^2} \Big\vert
\sum_{n=1}^{\infty}
\les(-)^n(2n+1) \left({a_n}-{b_n}\right)\ris
\Big\vert^2
\end{equation}
is the back-scattering efficiency, and
\begin{equation}
Q\forw= \frac{1}{4\pi^2\bara^2} \Big\vert
\sum_{n=1}^{\infty}
\les (2n+1) \left({a_n}+{b_n}\right)\ris
\Big\vert^2
\end{equation}
is
the forward-scattering efficiency 
 \cite{BH83,Lpr,Lnih}.
The coefficients 
\begin{equation}
\left.\begin{array}{l}
\label{eq53}
a_n=\displaystyle{
 \frac{\epsr\,j_n(2\pi\bara\nr)\psi_n\one(2\pi\bara)
-j_n(2\pi\bara)\psi_n\one(2\pi\bara\nr)}
{\epsr\,j_n(2\pi\bara\nr)\psi_n\three(2\pi\bara)
-h_n\one(2\pi\bara)\psi_n\one(2\pi\bara\nr)}
}
\\[10pt]
b_n= \displaystyle{
\frac{j_n(2\pi\bara\nr)\psi_n\one(2\pi\bara)
-j_n(2\pi\bara)\psi_n\one(2\pi\bara\nr)}
{j_n(2\pi\bara\nr)\psi_n\three(2\pi\bara)
-h_n\one(2\pi\bara)\psi_n\one(2\pi\bara\nr)} 
}
\end{array}\right\}\,,
\end{equation}
emerge from the solution of a boundary-value problem \cite{BH83},
with
\begin{equation}
\left.\begin{array}{l}
\nr = \sqrt{\epsr}
\\[5pt]
\psi_n\one(u) = \displaystyle{
\frac{d}{du}\les u\, j_n(u)\ris
}
\\[8pt]
\psi_n\three(u) = \displaystyle{
\frac{d}{du}\les u \,h_n\one(u)\ris
}
\end{array}
\right\}\,,
\end{equation}
where $j_n(u)$ is the spherical Bessel function of order $n$ and argument $u$
and $h_n\one(u)$ is the spherical Hankel function of the first kind, order $n$, and argument $u$.
An $\exp(-i\omega t)$ dependence on time $t$ is implicit, with $i=\sqrt{-1}$ and
$\omega$ denoting the angular frequency.

\begin{figure}[h]
\centering
\includegraphics[width=5cm]{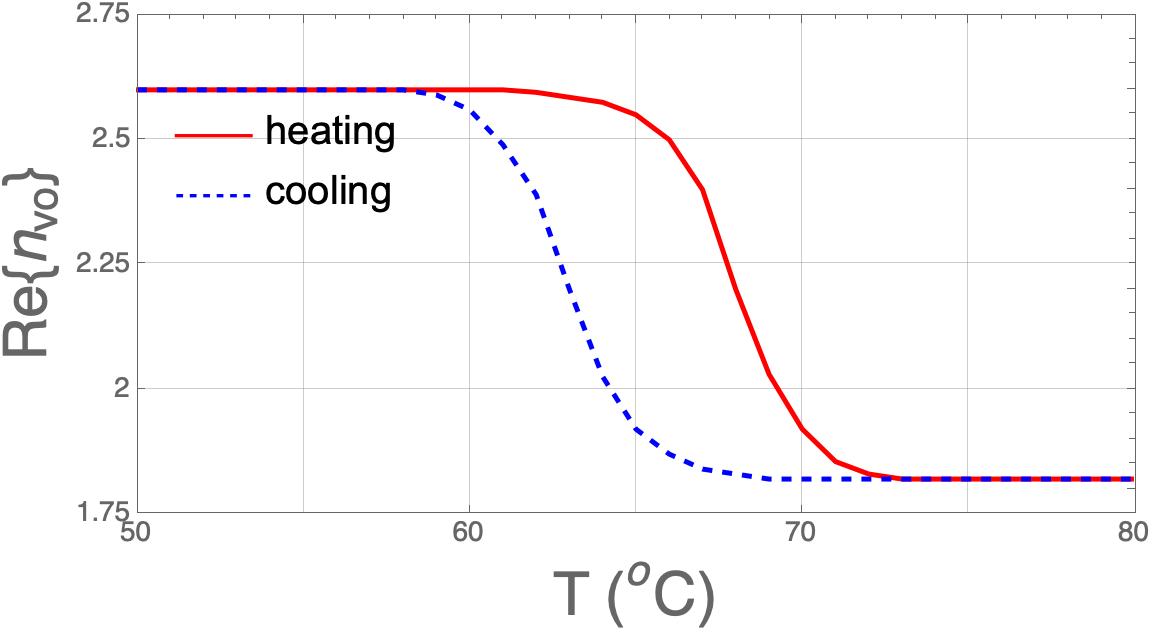} 
\includegraphics[width=5cm]{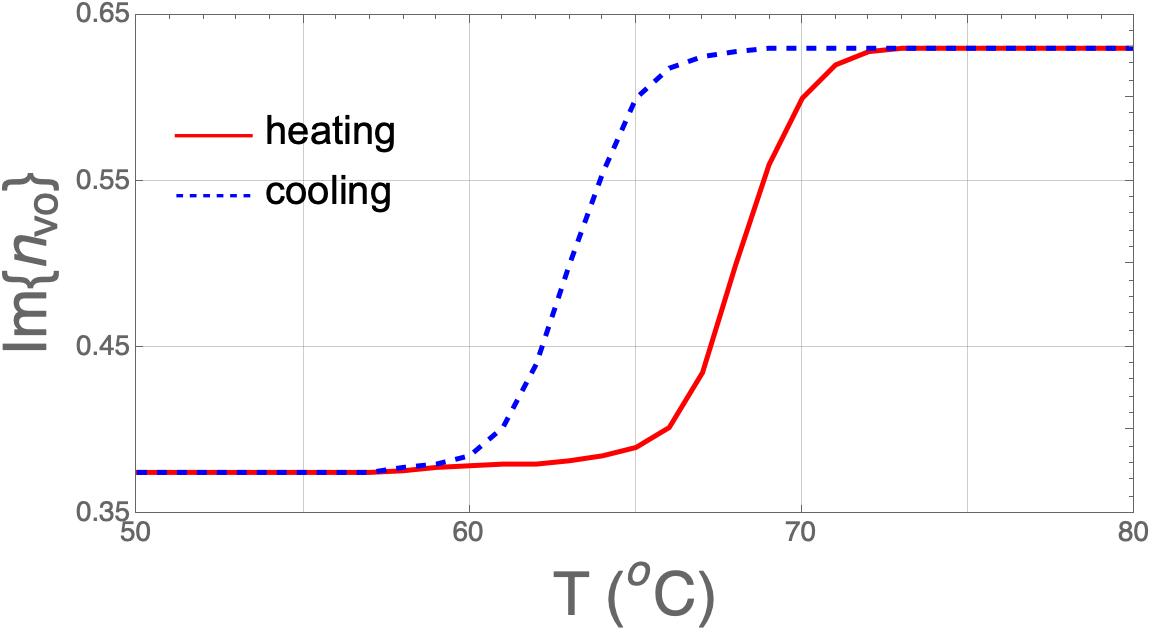}   
\caption{${\rm Re}\lec\nr\ric$ and ${\rm Im}\lec\nr\ric$ as functions of $T$ for $\lambdao=800$~nm
\cite{Cormier}. }
\label{Fig:n800}
\end{figure}

\begin{figure}[h]
\centering
\includegraphics[width=5cm]{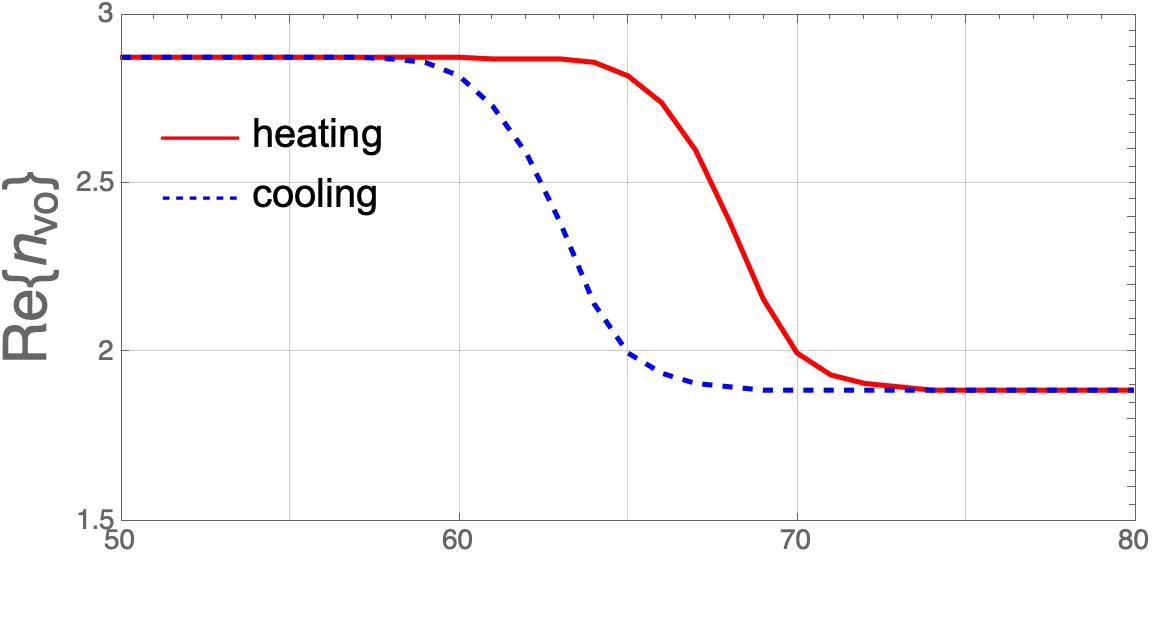} 
\includegraphics[width=5cm]{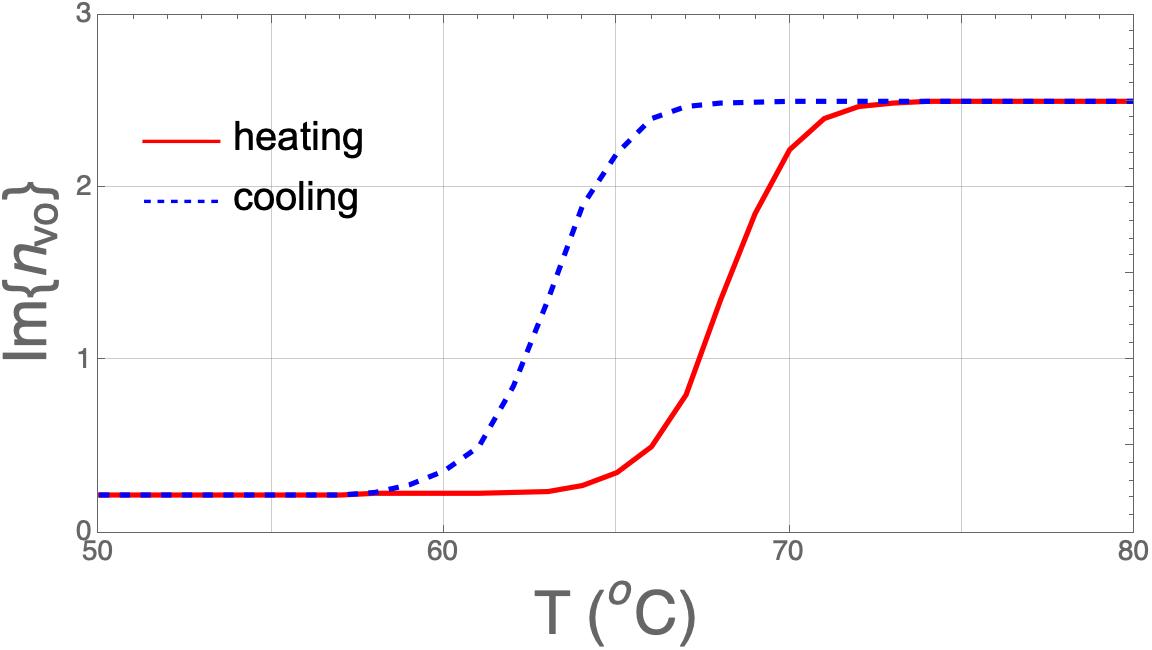}   
\caption{Same as Fig.~\ref{Fig:n800} except for $\lambdao=1550$~nm \cite{Cormier}. }
\label{Fig:n1550}
\end{figure}

\section{Numerical Results and Discussion}
Figures~\ref{Fig:n800} and \ref{Fig:n1550} present the real and
and imaginary parts of $\nr$   as
functions of $T$ for $\lambdao=800$~nm and $\lambdao=1550$~nm, respectively \cite{Cormier}.
Since ${\rm Re}\lec\nrm\ric\simeq 7\,{\rm Im}\lec\nrm\ric$ for $\lambdao=800$~nm
and ${\rm Re}\lec\nrm\ric\simeq 13\,{\rm Im}\lec\nrm\ric$ for $\lambdao=1550$~nm,
monoclinic \vo~is a moderately dissipative insulator for both wavelengths. 
However, ${\rm Re}\lec\nrt\ric\simeq 3\,{\rm Im}\lec\nrt\ric$ for $\lambdao=800$~nm
but ${\rm Re}\lec\nrt\ric\simeq 0.75\,{\rm Im}\lec\nrt\ric$ for $\lambdao=1550$~nm.
Accordingly,  tetragonal \vo~is
\begin{itemize}  
\item[(i)] a more dissipative insulator  than monoclinic \vo~for $\lambdao=800$~nm,
but
\item[(ii)] a plasmonic metal for   $\lambdao=1550$~nm.
\end{itemize}
Thus, 
\begin{itemize}  
\item[(i)]  both heating and cooling engender
an insulator-to-insulator transformation (IIT)
for $\lambdao=800$~nm because ${\rm Re}\lec\epsvom\ric>0$ and ${\rm Re}\lec\epsvot\ric>0$,
\end{itemize}
but
\begin{itemize}
\item[(ii)]  heating gives rise to the IMT
whereas cooling creates the  MIT
for $\lambdao=1550$~nm because ${\rm Re}\lec\epsvom\ric>0$ and ${\rm Re}\lec\epsvot\ric<0$.
\end{itemize}
Switching applications of \vo~exploit the   IMT and MIT \cite{Dumas,Wang2015,Hashemi,Sereb2022,Tripathi}
but ignore the IIT and avoid the intermediate thermal regime.
 
\newpage

\begin{widetext}
\begin{figure}[h]
\centering
\includegraphics[width=11cm]{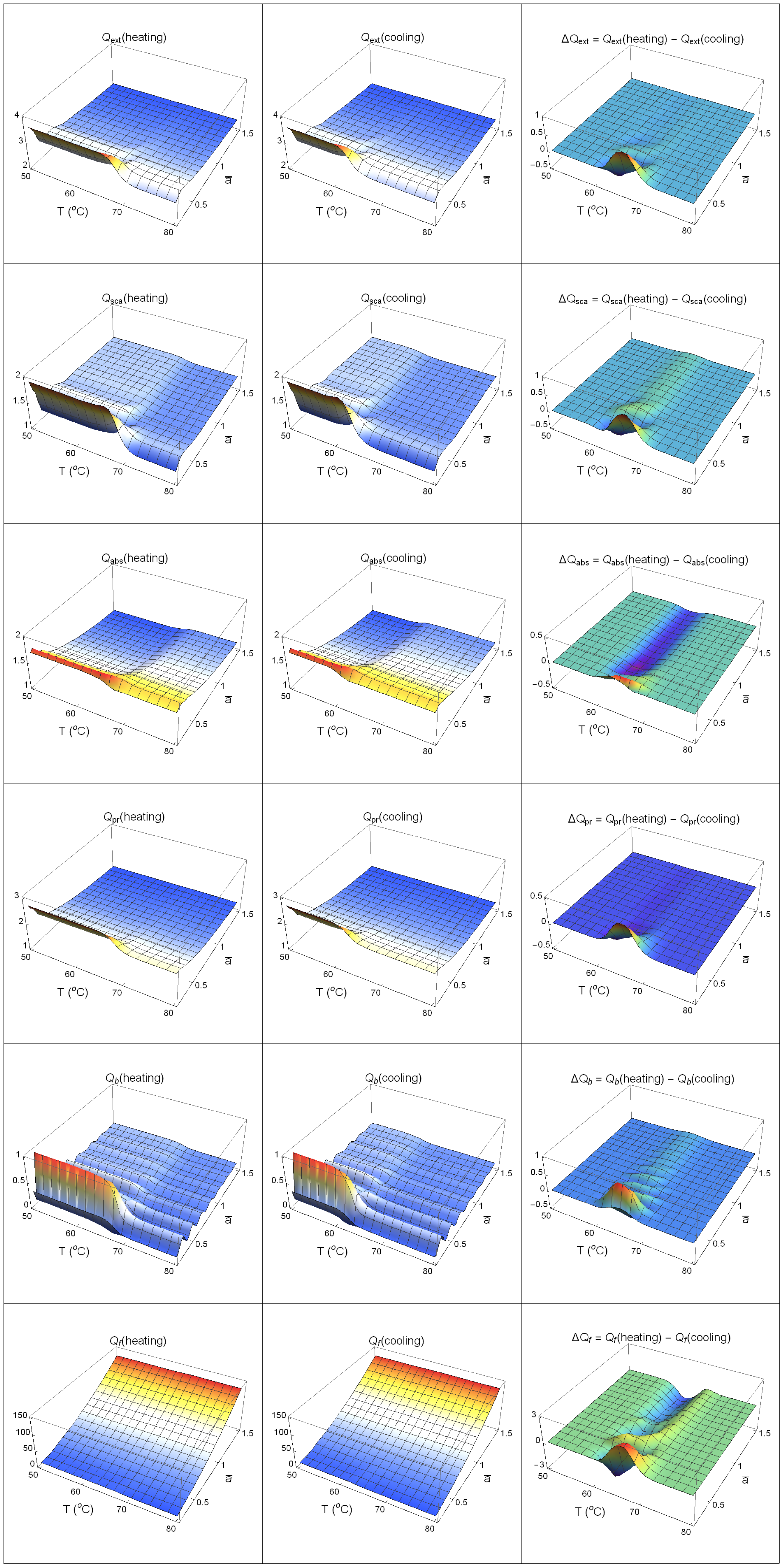} 
\caption{$Q\ext$, $Q\sca$, $Q\abs$, $Q\pr$, $Q\back$, and $Q\forw$ as functions of $T$
and $\bara$ for $\lambdao=800$~nm. }
\label{Fig:Q800}
\end{figure}
\end{widetext}

\begin{widetext}
\begin{figure}[h]
\centering
\includegraphics[width=11cm]{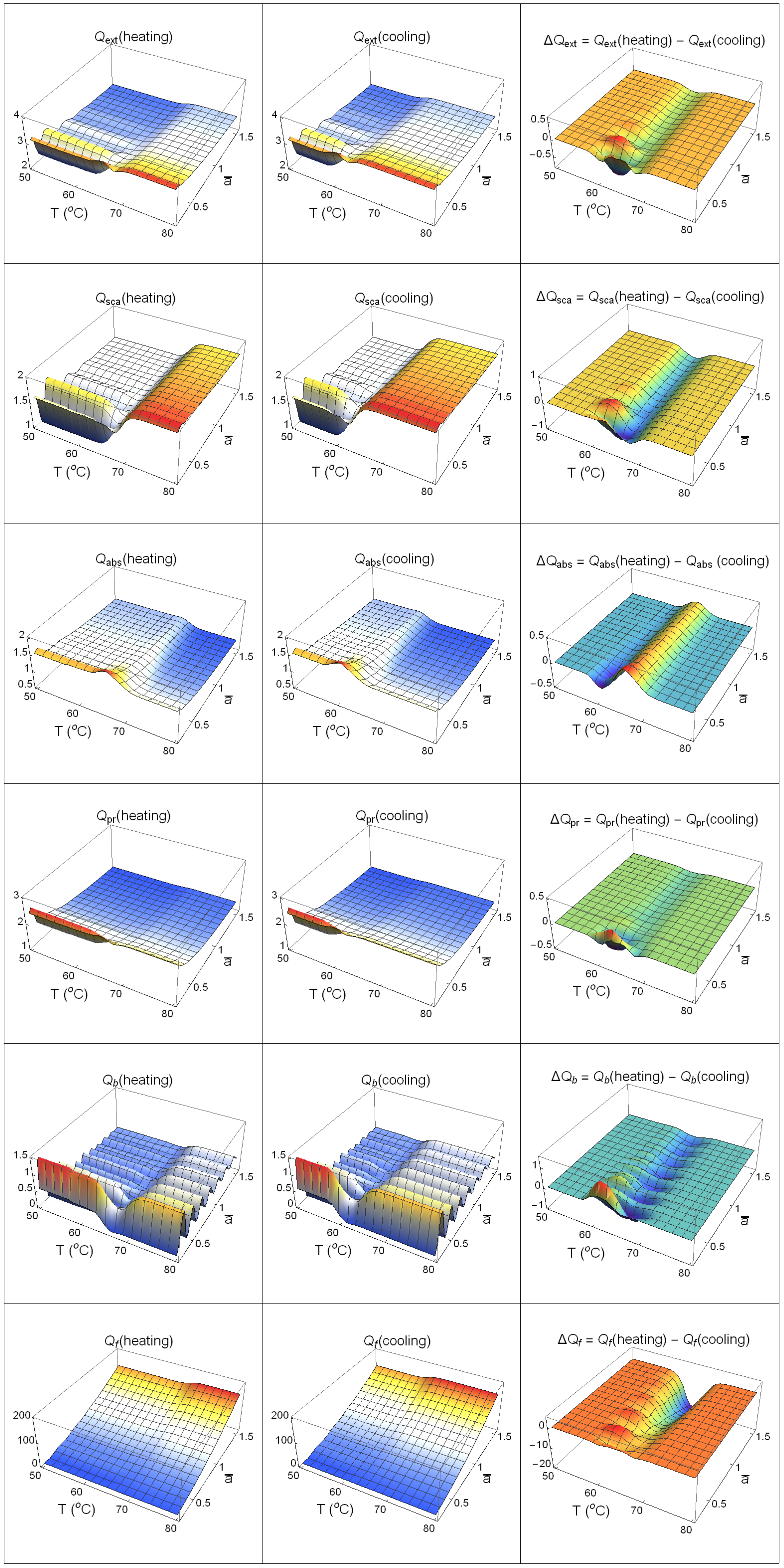} 
\caption{Same as Fig.~\ref{Fig:Q800} except for $\lambdao=1550$~nm. }
\label{Fig:Q1550}
\end{figure}
\end{widetext}

\subsection{IIT/IIT ($\lambdao<1100$~nm)}
Both heating and cooling engender
a reversible IIT for $\lambdao=800$~nm, as is clear from Fig.~\ref{Fig:n800}. Figure~\ref{Fig:Q800}
provides graphs of $Q\ext$, $Q\sca$, $Q\abs$,   $Q\pr$, $Q\back$, and $Q\forw$ as functions of
$T$ and $\bara$ on both the heating and the cooling branches. The first four efficiencies in that list of six
may be considered as \textit{directionally averaged} metrics of scattering, whereas   $Q\back$ refers
to the back-scattering direction
and $Q\forw$ to the forward-scattering direction.
Furthermore, the difference 
\begin{eqnarray}
\nonumber
&&
\DQ_{\ell}=Q_{\ell}{\rm (heating)}-Q_{\ell}{\rm (cooling)}\,, 
\\[5pt]
&&\qquad\qquad
\ell\in\lec{\rm ext},\,{\rm sca},\,{\rm abs},\,{\rm pr},\,{\rm b},\,{\rm f}\ric\,,
\end{eqnarray}
of each of the six efficiencies on the heating branch relative to its value on the cooling branch
is also plotted with respect to $T$ and $\bara$ in Fig.~\ref{Fig:Q800}; obviously, this difference does not exist
for $T\lesssim58$~\degC~and $T\gtrsim72$~\degC.

All six efficiencies in Fig.~\ref{Fig:Q800} are different  for purely monoclinic \vo~(i.e., $T\lesssim58$~\degC) and purely
tetragonal \vo~(i.e., $T\gtrsim72$~\degC). 
Unsurprisingly therefore,    significant dependences on temperature exist in
the  intermediate thermal regime on both the heating and cooling branches.

The graphs of $Q\ext$ and $\DQ\ext$ indicate that the effect of thermal hysteresis on plane-wave extinction
 is much more pronounced
 when the radius of the \vo~sphere
is less than half the free-space wavelength, with $\DQ\ext$ being of the highest magnitude somewhere
in the middle third of the intermediate thermal regime.  As $\bara$ increases from $0.5$, the effect of thermal
hysteresis diminishes rapidly and is barely noticeable when $\bara=1.5$. This may be related to the
fact that $Q\ext\to2$ as $\bara\to\infty$, regardless of the composition of the sphere \cite{Brillouin,Berg}.

The effects of thermal hysteresis on  $Q\sca$ and $Q\abs$ 
are much more pronounced
 when $\bara\lesssim0.5$.
Both $\DQ\sca$ and $\DQ\abs$ are also of the highest magnitude somewhere
in the middle third of the intermediate thermal regime. Their magnitudes diminish  as $\bara$ increases beyond
about $0.5$, with $\DQ\sca$ and $\DQ\abs$ roughly equal in magnitude but opposite in sign, which is
not surprising since
$\DQ\ext=\DQ\sca+\DQ\abs$.

The graphs of $Q\pr$ and $\DQ\pr$ are similar to those of $Q\ext$ and $\DQ\ext$, respectively. Hence,
the effect of thermal hysteresis on radiation-pressure efficiency
 is much more pronounced for $\bara\lesssim0.5$ than for $\bara\gtrsim0.5$.
 The effect of the IIT  is barely noticeable for any $\bara\gtrsim1$, i.e., $Q\pr$ depends very weakly
 on $T$ if \vo~is purely monoclinic or purely tetragonal, or if it comprises crystals of both types in any
 proportion. 
 
 The back-scattering efficiency is an undulating function of $\bara$ at any $T\in[50\,^\circ{\rm C}, 80\,^\circ{\rm C}]$,
 the undulations diminishing in magnitude as $\bara$ increases beyond $0.5$. In general, $Q\back$ is higher
 for purely monoclinic \vo~than for purely tetragonal \vo. The graph of $\DQ\back$ indicates that thermal hysteresis
 is most pronounced at $\bara\sim 0.35$.
 
 In contrast to $Q\back$, $Q\forw$ increases monotonically with $\bara\in[0.25,1.5]$ in Fig.~\ref{Fig:Q800} for 
  any $T\in[50\,^\circ{\rm C}, 80\,^\circ{\rm C}]$. The difference $\DQ\forw$ can be negative or positive as $T$
  changes in the intermediate thermal regime for $\bara\lesssim 1$. However, as $\bara$ increases beyond unity,
  $\DQ\forw$ becomes more negative.
  
  The overall conclusions from Fig.~\ref{Fig:Q800} are as follows:
  \begin{itemize}
  \item $Q\ext$, $Q\sca$, $Q\abs$,   $Q\pr$, and $Q\back$ are higher for $T\lesssim 58$~\degC~(purely monoclinic)
  than for $T\gtrsim 72$~\degC~(purely tetragonal), when the sphere is small ($\bara\lesssim0.5$).
  
  \item $Q\ext$, $Q\sca$, $Q\abs$,   $Q\pr$, and $Q\back$ are marginally different for $T\lesssim 58$~\degC~(purely monoclinic)
  from their values for $T\gtrsim 72$~\degC~(purely tetragonal), when the sphere is large ($\bara\gtrsim0.5$).
  
  \item All six efficiencies are affected by thermal hysteresis most when the sphere is small ($\bara\lesssim0.5$).
  
  \item $Q\ext$ and $Q\pr$ are affected very little by thermal hysteresis when the sphere is large ($\bara\gtrsim0.5$).
  
  \item $Q\sca$, $Q\abs$, $Q\back$, and $Q\forw$ are affected steadily by thermal hysteresis as $\bara$ increases beyond unity.

  \end{itemize}

\subsection{IMT/MIT ($\lambdao>1100$~nm)}

When the free-space wavelength  increases
beyond 1100~nm, tetragonal \vo~becomes a plasmonic metal whereas monoclinic \vo~remains a dissipative insulator. 
The IMT/MIT is therefore very different from the IIT, as even a casual comparison of Figs.~\ref{Fig:Q800} (for $\lambdao=800$~nm)
and  \ref{Fig:Q1550} (for $\lambdao=1550$~nm) shows. Of course, every one of the six efficiencies in Fig.~\ref{Fig:Q1550} is different  for purely monoclinic 
\vo~than for purely
tetragonal \vo, and there are  significant dependences on temperature   in
the  intermediate thermal regime on both the heating and cooling branches.

The graph of $Q\ext$ in Fig.~\ref{Fig:Q1550} shows that the extinction  is more when \vo~is a metal (tetragonal)
rather than a dissipative dielectric (monoclinic), which is also in accord with Fig.~\ref{Fig:Q800}. Furthermore,
the graphs of $Q\ext$ 
and $\DQ\ext$ in Fig.~\ref{Fig:Q1550} indicate that the effect of thermal hysteresis on plane-wave extinction
 is somewhat more pronounced for $\bara\lesssim0.5$ than for $\bara\gtrsim0.5$. Whereas $\DQ\ext$ can assume both negative
 and positive values for $\bara\lesssim0.5$ in the intermediate thermal regime, it is predominantly negative for $\bara\gtrsim0.5$
 in the same regime. Certainly,  as $\bara\to\infty$ $Q\ext$ must go to $2$ \cite{Brillouin,Berg}, but   at a considerably slower rate
 with the increase in $\bara$ than in Fig.~\ref{Fig:Q800}. 
 
 Also, whereas $\DQ\sca$ has a bump and $\DQ\abs$ has a trough  that roughly compensate each
 other for  $Q\ext$ to be barely affected by thermal hysteresis for $\bara\gtrsim0.5$   in Fig.~\ref{Fig:Q800},
 $\DQ\sca$ has a trough and $\DQ\abs$ has a bump that do not compensate each other
 so that $Q\ext$ is affected by thermal hysteresis  for $\bara\gtrsim0.5$   in Fig.~\ref{Fig:Q1550}.
  
 Unlike in Fig.~\ref{Fig:Q800},
 the graphs of $Q\pr$  in Fig.~\ref{Fig:Q1550} are qualitatively different from   those of $Q\ext$. 
 The signature of thermal hysteresis for  $\bara\gtrsim0.5$ is somewhat more marked when
 IMT/MIT occur than when IIT occurs.
 
 The back-scattering efficiency is an undulating function of $\bara$ at any $T\in[50\,^\circ{\rm C}, 80\,^\circ{\rm C}]$,
 the undulations diminishing in magnitude as $\bara$ increases beyond $0.5$. In general, $Q\back$ is lower
 for purely monoclinic \vo~than for purely tetragonal \vo, contrary to Fig.~\ref{Fig:Q800}. The graph of $\DQ\back$ indicates that thermal hysteresis
 is most pronounced at $\bara\sim 0.3$.

The forward-scattering efficiency increases monotonically with $\bara\in[0.25,1.5]$ in Fig.~\ref{Fig:Q1550} for 
  any $T\in[50\,^\circ{\rm C}, 80\,^\circ{\rm C}]$,  in contrast to $Q\back$. The same tendency is evident
  in Fig.~\ref{Fig:Q800}. However, $Q\forw$ is higher when tetragonal \vo~is a plasmonic metal (Fig.~\ref{Fig:Q1550})
  than when it is a dissipative dielectric (Fig.~\ref{Fig:Q800}). Also,   $\DQ\forw$ is more noticeably negative in the intermediate thermal regime
  as $\bara$ increases beyond unity when the IMT/MIT  (Fig.~\ref{Fig:Q1550}) occur than when the IIT occurs  (Fig.~\ref{Fig:Q800}).
 
  The overall conclusions from Fig.~\ref{Fig:Q1550} are as follows:
  \begin{itemize}
  
  \item $Q\ext$, $Q\sca$,   $Q\back$, and $Q\forw$ are noticeably lower for $T\lesssim 58$~\degC~(purely monoclinic)
  than for $T\gtrsim 72$~\degC~(purely tetragonal), when the sphere is small ($\bara\lesssim0.5$).  $Q\sca$ shows
  the opposite trend, whereas $Q\pr$ is about the same for both  $T\lesssim 58$~\degC~and  $T\gtrsim 72$~\degC.

  \item All six efficiencies are   affected by thermal hysteresis, regardless of the sphere radius.

  \item $Q\ext$, $Q\sca$, $Q\abs$, $Q\pr$, and $Q\back$  are affected steadily by thermal hysteresis as $\bara$ increases beyond unity.
  
 \item $Q\ext$, $Q\sca$, $Q\abs$, $Q\back$, and $Q\forw$ are affected more in the intermediate thermal regime
 when tetragonal \vo~is a plasmonic metal (Fig.~\ref{Fig:Q1550}) than a dissipative dielectric (Fig.~\ref{Fig:Q800}). 

\end{itemize}

\begin{figure}[h]
\centering
\includegraphics[width=8cm]{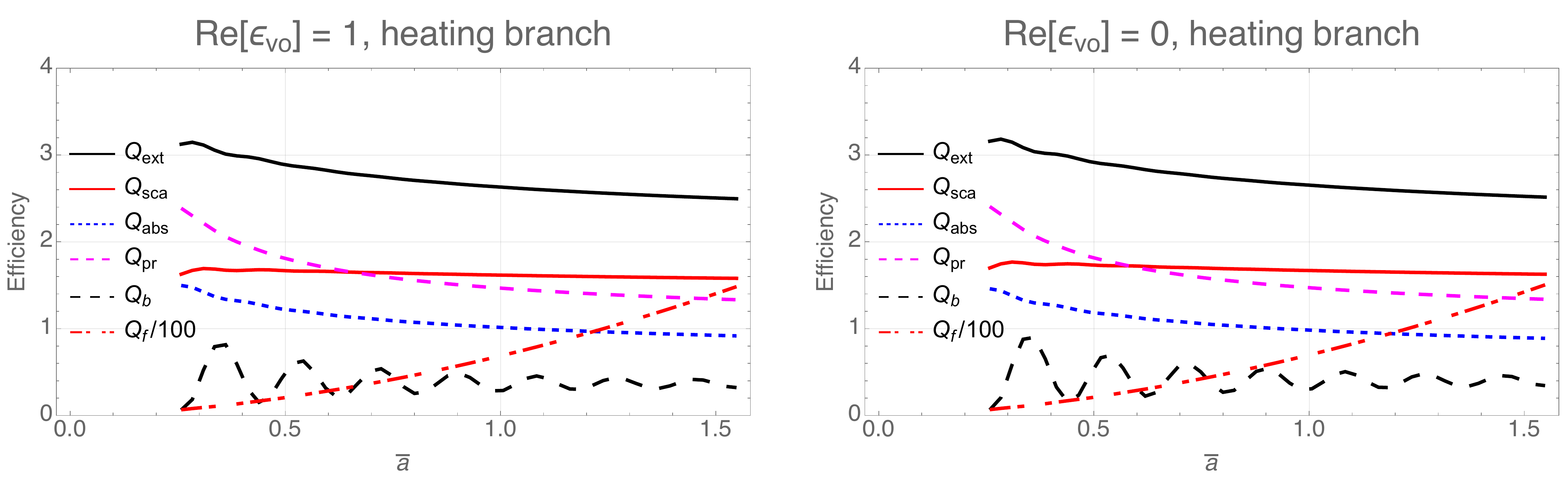} 
\caption{
$Q\ext$, $Q\sca$, $Q\abs$, $Q\pr$, $Q\back$, and $Q\forw$ as functions of $\bara$ 
on the heating branch when $\lambdao=1550$~nm. Left: $T=69.11$~\degC~(${\rm Re}\lec\epsr\ric=1+8.09i$).
Right: $T=69.57$~\degC~(${\rm Re}\lec\epsr\ric=8.50i$).}
\label{Fig:sp}
\end{figure}

\subsection{Vacuum and null-permittivity quasistates}
The IMT/MIT  also serves as a route to two  electromagnetic quasistates of matter. The first is
the \textit{vacuum quasistate} \cite{Sereb2018}: ${\rm Re}\lec\epsr\ric=1$. The second is the \textit{null-permittivity quasistate} \cite{Niu}:
${\rm Re}\lec\epsr\ric=0$. Both quasistates are exhibited by many materials that undergo temperature-mediated IMT/MIT, a very good example being indium
antimonide (InSb) \cite{ML-IEEEPJ,Alkhoori2022}. However, the IMT/MIT exhibited by InSb is not hysteretic, in contrast to \vo.

For $\lambdao=1550$~nm, we estimated from Fig.~\ref{Fig:n1550} that
\begin{itemize}
\item $\epsr$ =$1+8.09i$ at $T=69.11$~\degC~and
\item $\epsr$ = $8.50i$ at $T=69.57$~\degC
\end{itemize}
on the heating branch, whereas
\begin{itemize}
\item $\epsr$ =$1+8.13i$ at $T=64.01$~\degC~and
\item $\epsr$ = $8.50i$ at $T=64.55$~\degC
\end{itemize}
on the cooling branch. Figure~\ref{Fig:sp} shows all six efficiencies as functions of $\bara$   when \vo~is in the
vacuum and null-permittivity quasistates on the heating branch. Since the values of ${\rm Im}\lec\epsr\ric$ are roughly  the same  in both
quasistates and considerably exceed unity, there is practically no distinction in the efficiencies
for the two quasistates. As  similar findings apply on the cooling branch, there is no need to present the corresponding graphs of the six efficiencies.

\section{Concluding Remarks}
\label{conc}

As temperature changes from about 58~\degC~to 72~\degC, the crystal
structure of \vo~transforms from purely monoclinic to purely tetragonal, the transformation being reversible
but hysteretic.
We have studied the response characteristics of a \vo~sphere to planewave illumination to examine the effects of
thermal hysteresis on the extinction, total scattering, absorption, radiation-pressure, back-scattering, and forward-scattering 
 efficiencies.
 
Whereas monoclinic \vo~is a dissipative dielectric, tetragonal \vo~is a plasmonic metal for $\lambdao>1100$~nm but a dissipative
dielectric for  $\lambdao<1100$~nm. Thus, two distinct types of reversible but hysteretic transformations are possible. 
Clear signatures
of thermal hysteresis  in the insulator-to-insulator transformation are to be found in $Q\forw$, $Q\abs$, and $Q\back$ (in that order, in our
opinion). Clear signatures
of thermal hysteresis  in the insulator/metal-to-metal/insulator transformation are to be found in $Q\forw$, $Q\abs$, $Q\sca$,
and $Q\back$ (in that order). The contrast between the two dissipative dielectric forms of \vo~($\lambdao<1100$~nm) being less pronounced than
the contrast between the dissipative dielectric and the metallic forms of \vo~($\lambdao>1100$~nm), the IMT/MIT 
has clearer  signatures
of thermal hysteresis  than the IIT.

Thermal hysteresis is responsible for the double occurrences of the vacuum and null-permittivity quasistates in the intermediate
thermal regime when tetragonal \vo~is a plasmonic metal. However, none of the six efficiencies show significant differences
between the two quasistates of \vo.

The electromagnetic consequences of thermal hysteresis exhibited by \vo~may depend on the rate of heating or cooling,
but that dependence has turned out to be difficult to characterize \cite{Hache}. Photoexcitation by a free-electron laser has been
shown to effect the monoclinic-to-tetragonal transformation within 150~fs \cite{Wall}, but not the reverse transformation----which does require
cooling.
 
 \vspace{5mm}
\noindent {\bf Acknowledgments.} 
AL   thanks  the Charles Godfrey Binder Endowment at Penn State for ongoing support of his research efforts.
WIW thanks the Higher Education Commission of Pakistan
for an IRSIP Scholarship that enabled him to spend six months at    The Pennsylvania State University.

\end{document}